\documentclass[12pt]{JHEP3}
\usepackage{latexsym}
\usepackage{epsfig}

\newcommand{\eq}{\begin{equation}}
\newcommand{\feq}{\end{equation}}
\newcommand{\eqn}{\begin{eqnarray}}
\newcommand{\feqn}{\end{eqnarray}}
\newcommand{\arr}{\begin{eqnarray*}}
\newcommand{\farr}{\end{eqnarray*}}
\newcommand{\beq}{\begin{equation}}
\newcommand{\eeq}{\end{equation}}
\newcommand{\bea}{\begin{eqnarray}}
\newcommand{\eea}{\end{eqnarray}}

\def\beq{\begin{equation}}
\def\eeq{\end{equation}}
\def\bea{\begin{eqnarray}}
\def\eea{\end{eqnarray}}
\def\bc{\begin{displaymath}}
\def\ec{\end{displaymath}}
\def\lb{\label}

\def\ka{\kappa}
\def\la{\lambda}
\def\laq{\lambda^{2}}

\def\lb{\label}

\title{Tachyons in de Sitter Space and Analytical Continuation from
dS/CFT to AdS/CFT} 
\author{Mariano Cadoni and  Paolo Carta \\
Universit\`a degli Studi di Cagliari, Dipartimento di
Fisica and\\ INFN, Sezione di Cagliari, Cittadella Universitaria 09042,
Monserrato, Italy. \\
Email: \email{mariano.cadoni@ca.infn.it},
\email{paolo.carta@ca.infn.it}}
\abstract{We discuss analytic continuation from $d$ dimensional
Lorentzian de Sitter (dS$_d$) to $d$-dimensional Lorentzian anti-de
Sitter (AdS$_{d}$) spacetime.  We show that AdS$_{d}$, with opposite
signature of the metric, can be obtained as analytic continuation of a
portion of dS$_d$.  This implies that the dynamics of
(positive square-mass) scalar particles in AdS$_{d}$ can be obtained
from the dynamics of tachyons in dS$_d$. We discuss this
correspondence both at the level of the solution of the field
equations and of the Green functions.  The AdS/CFT duality is
obtained as analytic continuation of the dS/CFT duality.}
\preprint{INFNCA-TH0209}
\begin{document}
\section{Introduction}
One of the most important realization of the holographic principle
\cite{'tHooft:gx,Susskind:1994vu} is the correspondence between bulk
anti-de Sitter (AdS) gravity and boundary conformal field theories
(CFT) \cite{Maldacena:1997re,Gubser:1998bc,Witten:1998qj}.  The
success of the AdS/CFT correspondence has motivated further proposal
of holographic dualities between gravitational systems and conformal
field theories.  In particular, a similar  correspondence 
has been proposed  for gravity in de Sitter 
(dS) spacetime (dS/CFT) \cite{Strominger:2001pn} (see also for 
instance 
\cite{Witten:2001kn,Klemm:2001ea,Spradlin:2001pw,Spradlin:2001nb}).

$d$-dimensional de Sitter spacetime (dS$_d$) looks, at least
locally, very similar to $d$-dimensional anti-de Sitter spacetime
(AdS$_{d}$).  They are both spacetimes of constant curvature and
dS$_d$ can be obtained from AdS$_{d}$  by flipping the sign (from negative to
positive) of the cosmological constant $\Lambda$. Naively, one could
therefore expect the dS$_{d}$/CFT$_{d-1}$ duality to be simply
related to AdS$_{d}$/CFT$_{d-1}$, for instance by means of a 
analytical continuation.  However, local and global properties of de
Sitter spacetime lead to unexpected obstructions.  AdS$_{d}$ has a
simply connected boundary, which is (a conformal compactification of)
 $d-1$ dimensional Minkowski space.  Conversely, dS$_d$ has two
disconnected boundaries conformally related to $d-1$ dimensional
Euclidean space.  Moreover, the causal structure of dS$_d$ is
completely different from that of AdS$_{d}$. A single de Sitter
observer has not access to the whole of dS$_d$. This has a strong
impact on the features of  quantum field theories on dS$_d$. It
is essentially responsible for the existence of a family of vacua
\cite{Allen:ux,Bousso:2001mw} for quantum scalar fields in dS$_d$,
whereas the corresponding AdS$_d$ vacuum is essentially unique.  Last
but not least, dS$_d$ is a time-dependent gravitational background,
which is very poorly understood in the context of string theory.

Because of these difficulties, the status of the dS/CFT correspondence
remains unclear and plugged by unresolved problems, which raised
criticisms about the very existence of a dS/CFT duality \cite{Dyson:2002nt,
Dyson:2002pf}.
A possible strategy to tackle the problems is to explore in detail the
similarities between  dS/CFT and AdS/CFT. After all dS$_d$ is
formally very similar to AdS$_{d}$, so that one can hope to find some
analytical continuation relating dS/CFT to AdS/CFT.  This approach has
been used in the Euclidean context. It has been shown that
AdS$_{d}$ can be considered as ``negative'' Euclideanization of
dS$_d$ \cite{McInnes:2001dq}.  However, the Euclidean approach met
only partial success, essentially because to achieve real progress one
needs not only to analytically  continue  the spacetime metric but also
the Green functions. In the Euclidean
context, no analytic continuation relating the AdS vacuum with some of
the dS vacua could be found \cite{Bousso:2001mw}.

In our opinion the dS/CFT duality can be understood as an analytical
continuation of AdS/CFT in the {\sl Lorentzian} context.  An essential
ingredient is represented by a correspondence between  tachyons in
dS$_{d}$ and  particles with positive square-mass in AdS$_{d}$.  In
Ref. \cite{McInnes:2001dq} it was argued that a  scalar particle
with positive square-mass
in dS space is like a tachyon in AdS space. Moreover, investigating 
the dS$_{2}$/CFT$_{1}$ correspondence
\cite{Cadoni:2002kz,Medved:2002tq} along the lines of the  
AdS$_{2}$/CFT$_{1}$
duality \cite{Cadoni:1998sg,Cadoni:1999ja}, it was shown that a
tachyonic perturbation in the two-dimensional bulk corresponds to a boundary
conformal operator of positive dimension.  The key idea we use in
this paper is very simple. The overall minus sign coming from the
analytical continuation of the metric is compensated by the minus sign
in front of the tachyon square-mass.

More in detail, in this paper we review the arguments of Ref. 
\cite{McInnes:2001dq},
based on unitarity bounds, about the correspondence between tachyons
in dS$_d$ and (positive square-mass) particles in AdS$_{d}$
(Sect. 2).  Using a particular coordinatization of dS$_d$, we show
that a region of Lorentzian dS$_d$ can be mapped, by means of a
analytic continuation, into whole ``negative'' Lorentzian
AdS$_{d}$. As a consequence, the scalar field equation for a tachyon in
dS$_d$ becomes that for a  scalar with positive square-mass 
in AdS$_{d}$ (Sect. 3).  
The particular chart used to cover dS$_d$ and the related causal
structure of the region of dS$_d$ mapped into AdS$_{d}$ allow for a
consistent mapping of the two spacetimes (Section 4).  We also show
explicitly that the solutions of the field equation and the Green 
functions for a tachyon in
dS$_d$ can be analytically continued into the solutions and Green
functions of a scalar field in AdS$_{d}$.  In particular, we argue that
the AdS vacuum can be obtained as analytical continuation of a
tachyonic dS vacuum (Sect. 5, 6).  Finally, the  AdS/CFT duality is
obtained as analytic continuation of the dS/CFT duality (Sect. 7).

\section{Conformal weights and unitarity bounds}

The dS$_{d}$/CFT$_{d-1}$ duality puts in correspondence a scalar field
of mass $m$ propagating in $d$-dimensional de Sitter space with a
$(d-1)$-dimensional CFT living on the boundary with conformal weights
\cite{Strominger:2001pn}
\beq\lb{e1}
h_{\pm}= {1\over 2}\left( d-1 
\pm\sqrt{(d-1)-{4m^{2}\over\la^{2}}}\,\right),
\eeq
where $\la^{2}$ is related to the cosmological constant $\Lambda$ of 
the de Sitter  space by $\Lambda= (d-2)(d-1)\la^{2}/2$ (In this paper 
we take $d>2$).

Conversely, the AdS$_{d}$/CFT$_{d-1}$ duality tells us that a scalar
field of mass $m$ propagating in $d$-dimensional anti-de Sitter space
is in correspondence with a $d$-dimensional boundary CFT with
conformal weight \cite{Witten:1998qj}
\beq\lb{e2}
\Delta= {1\over 2}\left( d-1 +\sqrt{(d-1)+{4m^{2}\over\la^{2}}}\,\right),
\eeq
now the spacetime has negative cosmological constant $-\Lambda$.  The
similarity between Eq. (\ref{e1}) and Eq. (\ref{e2}) is striking.  In
the case of the dS$_{d}$/CFT$_{d-1}$ correspondence the CFT is unitary
only if the Strominger bound, $m^{2}\le (d-1) (\la^{2}/4)$, for the mass
of the scalar is satisfied \cite{Strominger:2001pn}. For
AdS$_{d}$/CFT$_{d-1}$ the unitarity bound is given by the
Breitenlohner-Freedman bound, $m^{2}\ge - (d-1) (\la^{2}/4)$.  The two
bounds are related by the transformation $m^{2}\to -m^{2}$.  Moreover,
by changing the sign of $m^{2}$ in Eq. (\ref{e2}) we reproduce one of the
solutions in Eq. (\ref{e1}), $\Delta(-m^{2})=h_{+}(m^{2})$.  Thus,
scalar fields with positive $m^{2}$ in AdS$_{d}$ are like tachyons in
dS$_d$ and vice versa. This fact was first observed in Ref.
\cite{McInnes:2001dq}.

The previous statement represents just a conjecture, unless one can
map explicitly the dynamics of tachyons in dS$_{d}$ into that
of  scalar fields in AdS$_{d}$.  Moreover, one should
also be able to show explicitly that the transformation $m^{2}\to
-m^{2}$ maps the dS$_{d}$/CFT$_{d-1}$ correspondence into the
AdS$_{d}$/CFT$_{d-1}$ one.  This is not so easy, for at least two
reasons, already pointed out in the literature: (a) the topology of
dS$_d$ is completely different from that of AdS$_{d}$. In particular,
the boundary $\cal {B}$ of AdS$_{d}$ is simply connected and has the
topology of $S^{1}\times S^{d-2}$. dS$_d$ has two disconnected
boundaries, $\cal{I^{\pm}}$, each with topology $S^{d-1}$.  (b) until
now no analytic continuation between Lorentzian dS$_d$ and Lorentzian
AdS$_{d}$ could be found.

The previous features represent a strong obstruction that has to be
overcome if one wants to find a relationship between propagation 
of scalars in dS$_d$ and AdS$_{d}$ or, more in general, 
between dS$_{d}$/CFT$_{d-1}$ and
AdS$_{d}$/CFT$_{d-1}$.  For instance, the existence of the two
boundaries $\cal{I^{\pm}}$ for dS$_d$ implies that there are two
independent boundary conditions (corresponding to two independent sets
$\phi_{\pm}$ of boundary modes), which can be imposed on the
asymptotic behavior of the scalar field. Once the dS$_{d}$/CFT$_{d-1}$
correspondence is implemented, this implies the existence of  two
roots $h_{\pm}$ in Eq.  (\ref{e1}) for the conformal weights of the
boundary CFT, whereas in the AdS$_{d}$/CFT$_{d-1}$ only the single root
$\Delta$ of Eq.  (\ref{e2}) is present.  Also, the analytic
continuation from AdS$_{d}$ to dS$_d$ is a rather involved
problem. Owing to the different topologies, one cannot switch from
AdS$_{d}$ to dS$_d$ just by using the analytic continuation $\la\to i
\la$. A possibility to circumvent the problems is to work in the
Euclidean rather then in the Lorentzian.  Euclidean AdS$_{d}$ can be
considered as ``negative'' Euclideanization of dS$_d$
\cite{McInnes:2001dq}.  However, for obvious reasons the Euclidean
formulation cannot be used if one wants to show the existence of a
correspondence between tachyons on dS$_d$ and  particles 
with positive square-mass on
AdS$_d$. 

In the next section we will show that Lorentzian dS$_d$ can be
analytically continued into ``negative'' Lorentzian AdS$_{d}$, i.e. usual
anti-de Sitter spacetime with a metric tensor of opposite signature.
In the following we will only consider tachyons on dS$_d$ 
and the corresponding  scalar particles in AdS$_{d}$.  In
this case one has $h_{+}>0$ and $h_-<0$. However, our discussion could be
easily generalized to scalar fields in dS$_{d}$ satisfying the Strominger
bound and corresponding tachyons in AdS$_{d}$ satisfying the
Breitenlohner-Freedman bound.

\section{Analytic continuation from dS$_{d}$ to negative AdS$_{d}$} 
$d$-dimensional de Sitter spacetime can be defined as the hyperboloid
\beq\lb{hyp}
\eta_{AB}X^{A}X^{B}={1\over \laq},\quad \eta_{AB}=(-1,\ldots 1),\quad
A,B=0,1\ldots d,
\feq
embedded in the $d+1$-dimensional Minkowski spacetime,
$ds^{2}=\eta_{AB}dX^{A}dX^{B}$. The parametrization of the 
hyperboloid  suitable for our analytic continuation has been given 
in Ref. \cite{Klemm:2001ea}
\bea\lb{para}
\la X^{d}&=&\sqrt{\laq r^{2}+1}\sin\la t,
\quad \la X^{d-1}=\sqrt{\laq r^{2}+1}\cos\la t \nonumber\\
X^{0}&=&r \cosh \theta,\quad X^{j}= r \omega^{j} \sinh\theta,\, 
j=(1\ldots d-2),
\eea
with $r>0$, $0\le t\le 2\pi/\la$, $\theta>0$ and $\omega^{j}$ 
parametrizing 
the $S^{d-3}$ sphere, $ \omega^{1}= \cos\theta_{1},\, \omega^{2}= \sin\theta_{1}
\cos\theta_{2},\ldots,\,  \omega^{d-2}= 
\sin\theta_{1}\ldots\sin\theta_{d-3}$.
The induced metric on dS$_d$ is
\beq\lb{dsme}
ds^{2}=-(1+\laq r^{2})^{-1}dr^{2}+(1+\laq r^{2})dt^{2}
+r^{2}\left(d\theta^{2}+\sinh^{2}\theta d\Omega^{2}_{d-3}\right),
\eeq
where $d\Omega^{2}_{d-3}$ is the metric over the $S^{d-3}$ sphere.
Notice that the co\-or\-di\-nates $(\theta, \theta_{k})$ parametrize a
hyperbolic space $H^{d-2}$ with metric 
\beq\lb{hme}
d\Sigma^{2}=d\theta^{2}+
\sinh^{2}\theta d\Omega^{2}_{d-3}. 
\feq
The $r=const$ sections of the 
metric have  
$S^{1}\times H^{d-2}$ topology. The coordinatization (\ref{para}) 
gives an hyperbolic slicing of dS$_d$ and do 
not cover the whole de Sitter hyperboloid, but just a region of it.
In particular, only one boundary (for instance $\cal{I^{+}}$, which 
can be reached 
letting $r\to\infty$) of dS$_d$ is visible in these coordinates.
Moreover, using this coordinate system  $\cal{I^{+}}$ has 
$S^{1}\times H^{d-2}$ topology. Later on this paper, when discussing 
the causal structure of the spacetime, we will come back to 
this point.

We can now obtain negative AdS$_{d}$ just by using the analytic 
continuation 
\beq\lb{ac}
\theta\to i \theta,
\feq
in the de Sitter metric (\ref{dsme}).
We get 
\beq\lb{adsme}
ds^{2}=-\left[-(1+\laq r^{2})dt^{2}+(1+\laq r^{2})^{-1}dr^{2}
+r^{2}d\Omega^{2}_{d-2}\right],
\eeq
where now $d\Omega^{2}_{d-2}$ is the metric over the  $S^{d-2}$ sphere,
and the hyperbolic coordinate $\theta$ has become after the analytic 
continuation an angular coordinate of the sphere. 

The metric (\ref{adsme}) is (minus) the metric that can be used over
the whole AdS hyperboloid.  The AdS spacetime with metric given by
Eq. (\ref{adsme}) can be defined as the hyperboloid
\beq\lb{hyper}
(X^{d})^{2}+(X^{d-1})^{2}-\sum_{k=0}^{d-2}(X^{k})^{2}={1\over \la^{2}} \eeq
embedded in $d+1$-dimensional flat space with metric \beq\lb{embe}
ds^{2}=
-\left(-(dX^{d})^{2}-(dX^{d-1})^{2}+\sum_{k=0}^{d-2}(dX^{k})^{2}\right).
\eeq Notice that the signature of the embedding space is opposite to
the usual one used to define the AdS spacetime. The AdS hyperboloid
(\ref{hyper}) and AdS embedding space (\ref{embe}) can be obtained as
the analytic continuation $X^{j}\to iX^{j},\, j=1\ldots (d-2)$ (which
is the transformation (\ref{ac}) written in terms of the coordinates
$X$) of the dS hyperboloid (\ref{hyp}) and the dS embedding space.

The analytic continuation (\ref{ac}) changes the topology of the
$r=const$ sections of the metric.  In the AdS case (\ref{adsme}) the
topology of the $r=const$ sections is $S^{1}\times S^{d-2}$. The
analytic continuation $\theta\to i \theta$ maps $H^{d-2}\to S^{d-2}$.
It transforms the $\cal{I^{+}}$ boundary of dS$_d$ with topology
$S^{1}\times H^{d-2}$ into the boundary $\cal{B}$ of AdS$_{d}$ with topology
$S^{1}\times S^{d-2}$.

The metric (\ref{adsme}) corresponds to a parametrization of AdS space 
that covers the whole hyperboloid. In the following we will also 
make use of (negative) AdS metric written in terms of the Poincar\'e 
coordinates (see e.g. \cite{Aharony:1999ti})
\beq\lb{me}
ds^{2}=-\left[ {du^{2}\over\laq u^{2}}+ u^{2}\left(-d\tau^{2}+ 
dx_{j}dx^{j}\right)\right],
\feq
where $0\le u<\infty$ and $j=1\ldots(d-2)$.
These coordinates cover only half of the AdS hyperboloid.
The metric (\ref{me}) can be obtained as analytic continuation of the de 
Sitter metric in planar coordinates
\beq\lb{me1}
ds^{2}=-d\hat t^{2}+ e^{2\la \hat t} dx_{a}dx^{a}, \, a=1\ldots (d-1),
\feq
with $-\infty<\hat t<\infty$.
Performing in Eq. (\ref{me1}) the analytic continuation 
\beq\lb{ac1}
x_{j}\to 
ix_{j},\quad j=1,\ldots, (d-2),
\feq
and setting $e^{\la \hat t}=u$, $x_{d-1}=\tau$, we get 
immediately the metric (\ref{me}).

Let us now consider a tachyonic scalar  field $\phi$ with square-mass 
$\,-m^{2}$
propagating in the de Sitter spacetime (\ref{dsme}), the field 
equation reads
\beq\lb{fe}
(\nabla^{2}_{dS} +m^{2})\phi=0,
\feq
where the the operator $\nabla^{2}_{dS}$ is calculated with respect to the 
dS metric  (\ref{dsme}).
Performing in this equation the analytic continuation (\ref{ac}), we get the 
field  equation for  scalar field with a positive square-mass
propagating in the AdS spacetime,
\beq\lb{fe1}
(\nabla^{2}_{AdS} -m^{2})\phi=0,
\feq
where the the operator $\nabla^{2}_{AdS}$ is now calculated with respect to the 
usual  AdS metric, which is  given by Eq. (\ref{adsme}) with opposite 
sign. The minus sign coming from the analytic continuation is 
compensated by the change of sign of the square-mass of the scalar.

Of course the same holds true if we consider a tachyon propagating 
in dS$_{d}$ endowed  with the metric (\ref{me1}) and perform the analytic 
continuation  leading to the AdS metric in the form (\ref{me}).

We therefore conclude that, at least at the classical level, tachyonic
propagation in dS$_{d}$ is simply related by the analytic
continuation (\ref{ac}) to propagation of particles with
positive square-mass in AdS$_{d}$.

In our discussion, we did not take into consideration the global
structure of  both dS$_{d}$ and AdS$_{d}$.  We have already pointed out
in Sect. 2 that topologically AdS$_{d}$ and dS$_d$ are very
different. Therefore, our previous conclusion is strictly true only
locally.  In the next section we will discuss the impact that global
features of the spacetime have on the correspondence between tachyons
on dS$_d$ and particles with positive square-mass on AdS$_{d}$.

\section{Causal structure of the spacetime}
The coordinates $(r,t,\theta, \theta_{k})$ appearing in the hyperbolic 
slicing (\ref{dsme}) of dS$_d$ do not cover the full de Sitter 
hyperboloid but just a region of it. On the other hand, as it is 
evident from the metric (\ref{adsme}), the same coordinates
cover the whole AdS spacetime. This means that the analytic 
continuation (\ref{ac}) does not map  the whole dS$_d$  into  
entire AdS$_{d}$ but just a region of the dS hyperboloid into the 
the whole of the AdS hyperboloid.  In particular, only the $\cal{I^{+}}$ (i.e 
$r\to\infty$) boundary of dS$_d$ is visible in the particular 
coordinatization chosen, and it is mapped by means of the 
transformation (\ref{ac}) into the single  boundary $\cal{B}$ of 
AdS$_{d}$.  Although this fact does not have any influence on the local behavior
of the solution for the field equations (\ref{fe}), (\ref{fe1}),
the boundary condition we have to use when we solve these equations
depend crucially on the global features defined by the chart used to 
cover dS$_d$.

The region of dS$_d$ covered by the coordinates $(r,t,\theta,
\theta_{k})$ can be easily described by deriving the coordinate
transformation connecting them with the global coordinates $(\hat\tau,
\hat\theta_{a}), a=1,\ldots(d-1)$ that cover the whole of de Sitter
space. In terms of these coordinates the de Sitter metric reads
(see for instance Ref. \cite{Spradlin:2001pw} 
\beq\lb{gdsme} ds^{2}=-d\hat\tau^{2}+{1\over \la^{2}}\cosh^{2}(
\la\hat\tau) d\Omega^{2}_{d-1} , \feq
where $
-\infty<\hat\tau<\infty$ and $\hat\theta_{a}$ parametrize the
$S^{d-1}$ sphere whose metric is $d\Omega^{2}_{d-1}$. From the
coordinate transformation linking global and hyperbolic coordinates,
one easily finds that the coordinates $(r,t,\theta, \theta_{k})$ cover
only the region
\beq\lb{f1} \cosh^{2}(\la\hat \tau)
\sin^{2}\hat\theta_{1}\ldots \sin^{2}•\hat\theta_{d-2}\ge 1\,
\feq of dS$_d$ . The south and north
pole, $\hat\theta_{1},\ldots,\hat\theta_{d-2}=0,\pi$ are not covered
by the hyperbolic coordinate system $(r,t,\theta, \theta_{k})$.  This
feature explains why the topology of the $r=const$ (and in particular
of the boundary $r=\infty)$ sections of the metric (\ref{dsme}) is
$S^{1}\times H^{d-2}$ whereas that of the $\hat \tau=const$ (and in
particular of the boundaries $\cal{I^{\pm}}$) sections of the the
metric (\ref{gdsme}) is $S^{d-1}$.

\EPSFIGURE[h]{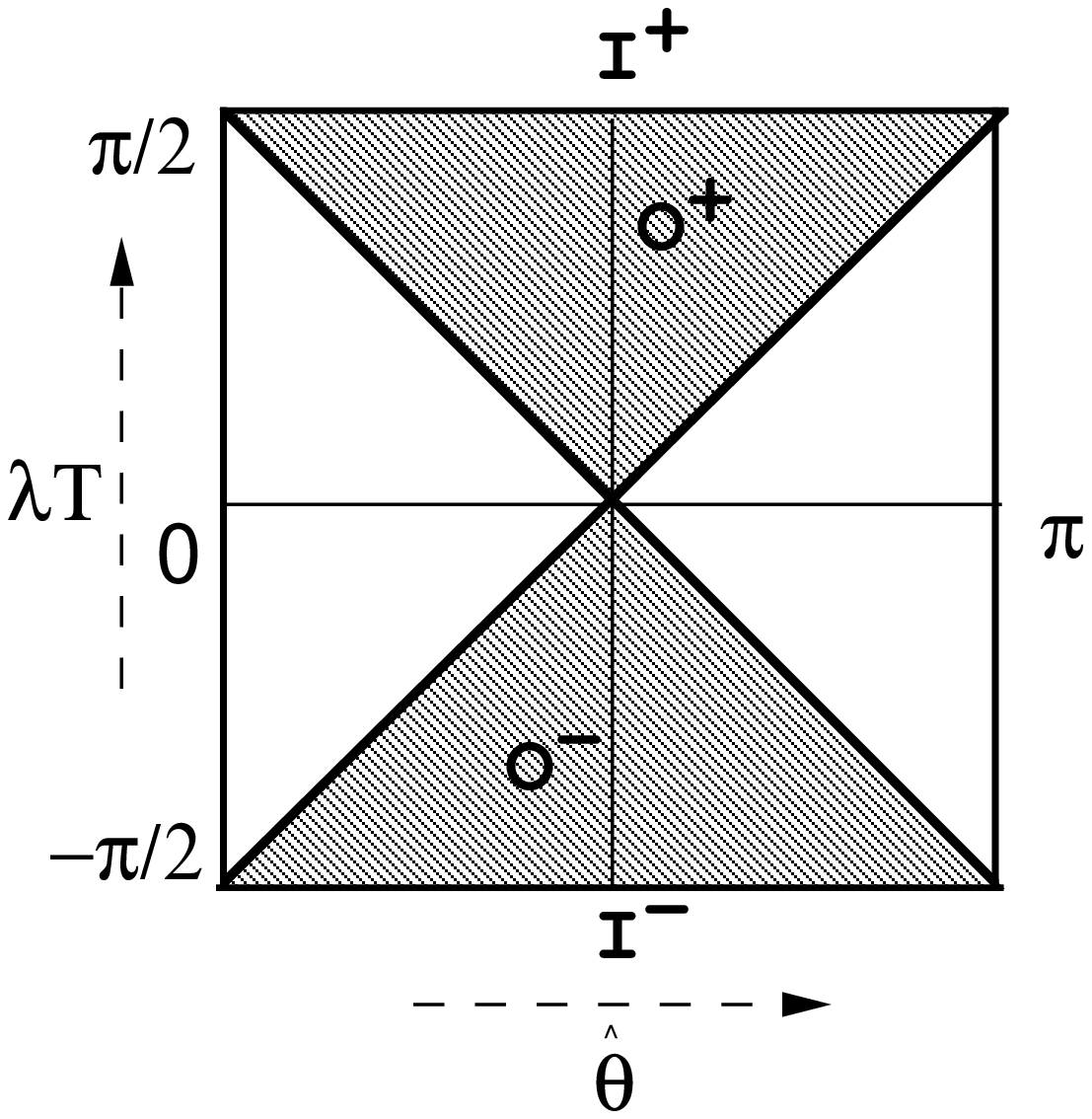,angle=0,width=7.5cm}{\label{frame} Penrose
diagram for dS$_d$} 
It is evident from Eq. (\ref{f1}) that given a point of the the
$S^{d-1}$ sphere with coordinates $(\hat\theta_{1}\ldots
\hat\theta_{d-1})$ its future, $\cal{O^{+}}$, (or past $\cal{O^{-}}$)
light-cone, is never completely covered by the hyperbolic coordinate
system unless Eq. (\ref{f1}) is identically satisfied. This is true
only for points with $\hat\theta_{1}\ldots \hat\theta_{d-2}=\pi/2$,
$0\le\hat\theta_{d-1}\le 2\pi$, i.e for the equator of the $S^{d-1}$
sphere. We conclude that the hyperbolic coordinates cover the 
region \- 
corresponding to the causal future (if the coordinate $r>0$) or the
causal past (if the coordinate $r<0$) of an observer  sitting at the
equator of the $S^{d-1}$ sphe\-re.  The Penrose diagram of the spacetime
can be easily constructed writing the metric (\ref{gdsme}) in
conformal coordinates, $(T, \hat\theta_{a})$, with $\cos(\la
T)=[\cos(\la \hat\tau)]^{-1}$, \beq\lb{gdsme1}
ds^{2}={1\over\cos^{2}(\la T)} \left(
-dT^{2}+\la^{-2}d\Omega^{2}_{d-1}\right) , \eeq where $ -\pi/2\le \la
T\le \pi/2$.  

Taking the angular coordinates of the sphere all
constant but one, $\hat\theta_{1}=\hat \theta$, we get the Penrose
diagram shown in figure 1.  The vertical lines $\hat
\theta=0,\pi/2,\pi$ represent the north pole, the equator and the
south pole, respectively. The horizontal lines ${\la T}=-\pi/2,\pi/2$
the two boundaries $\cal{I^{\pm}}$ of dS$_d$.  The region enclosed
between the two lines $\la T=\pm(-\hat\theta + {\pi\over2})$ is the region
covered by the hyperbolic coordinates. $\cal{O^{\pm}}$ are
respectively the future and past light-cone for an observer sitting at
$T=0$ at the equator. From our previous discussion emerges clearly
that if we restrict, as we did in Eq. (\ref{dsme}), the coordinate $r$
to range only over $[0,\infty]$ we cover only the region $\cal{O^{+}}$
of figure 1.  This means that $\cal{O^{+}}$ is the region of dS$_d$
which is mapped by the analytic continuation (\ref{ac}) into entire
AdS$_{d}$
\footnote {The choice of $\cal{O^{+}}$ instead of $\cal{O^{-}}$ is a
matter of conventions. We could have as well mapped $\cal{O^{-}}$ into
AdS$_{d}$.}.

Because $\cal{O^{+}}$ is the future light-cone of an
observer at the equator, we have reached the important result: the
portion of dS$_{d}$ corresponding to the causal future of
a single observer can be mapped by the simple analytic continuation
(\ref{ac})
into the whole AdS$_{d}$.  Translated in
terms of the dynamics of the scalar field $\phi$ considered in the
previous section: the dynamics of a tachyon defined in the region
$\cal{O^{+}}$ of dS$_{d}$, is analytically continued by Eq. (\ref{ac}) into the
dynamics of a  scalar field  (with positive square-mass) defined over 
the whole AdS$_{d}$ .

Let us conclude this section by observing that the correspondence 
between ${\cal O^{+}}\subset dS_{d}$  and AdS$_{d}$ is very natural from 
the physical point of view. It is well-known that a single observer 
cannot have access to the entire de Sitter spacetime.
Developing a field theory over dS$_d$ we should limit ourself to 
consider only regions over  which a single observer has a causal 
control (or from which can be causally controlled).
This point of view has been already advocated in Ref. 
\cite{Klemm:2001ea,Bousso:2001mw}. 

\section{Analytic continuation of the solutions for the scalar field}
In this section we will show explicitly that a solution of the field 
equation (\ref{fe}) for a tachyon in dS$_{d}$ can be analytically 
continued into a solution of the field equation (\ref{fe1}) for a 
scalar field in AdS$_{d}$.

Let us first consider the wave equation  (\ref{fe}) 
with the operator $\nabla^{2}$ evaluated on the metric (\ref{dsme}).
The differential equation (\ref{fe}) is separable.
The solution takes the form
\beq\lb{sol}
\phi(r,t,\Omega)= f(r)h(t)\hat Y_{d-2}(\Omega),
\feq
where $\Omega=(\theta, \theta_{1}\ldots\theta_{d-3})$.
$h(t)$  is solution of the equation
\beq\lb{sol1}
{d^{2}h\over dt^{2}}=-\laq n^2h,
\feq
with $n$ integer since $t$ is periodic.
$\hat Y_{d-2}(\Omega)$ are the  eigenfunctions of the $\nabla^2_{d-2}$ operator 
evaluated on the 
hyperbolic space $H^{d-2}$ with metric (\ref{hme}),
\beq\lb{sol2}
\nabla^2_{d-2}\hat Y_{d-2}=l(l+d-3)\hat Y_{d-2}.
\feq
$f(r)$ is  solution of the radial equation $('=d/dr)$
 \beq\lb{sol3}
f''+\left( {d-2\over r}+ {2r\laq \over (1+\laq r^2)}\right)f'+ 
\left[{n^2\laq\over (1+\laq r^2)^2} -{ l(l+d-3)\over r^2(1+\laq r^2)}-
{m^2\over (1+\laq r^2)}\right]f=0.
\feq
One can easily check that the wave equation (\ref{fe1}) for a scalar in 
AdS$_{d}$,
once the solution is separated as $\phi(r,t,\Omega)= f(r)h(t) Y_{d-2}(\Omega)$
with $\Omega=(\theta_{1}\ldots\theta_{d-2})$,  
exactly reproduces both Eq.  (\ref{sol1})  and   (\ref{sol3}) for 
$h(t)$ and $f(r)$, whereas
Eq.  (\ref{sol2}) is replaced by the equation
\beq\lb{sol2a}
\nabla^2_{d-2} Y_{d-2}=-l(l+d-3) Y_{d-2}.
\feq
where now the operator $\nabla^2_{d-2}$ is evaluated on the 
$S^{d-2}$ sphere.
$Y_{d-2}$ are the $(d-2)$-dimensional spherical harmonics.

We have still to show that the $(d-2)$-dimensional spherical harmonics
$Y_{d-2}$ are  analytical continuation of the
eigenfunctions $\hat Y_{d-2}$ of the Laplace-Beltrami operator in the
hyperbolic space $H^{d-2}$.  The Laplace-Beltrami operator has a
discrete spectrum in $S^{d-2}$ ($l$ in Eq. (\ref{sol2a}) must be a
nonnegative integer) but a continuous spectrum in $H^{d-2}$ ($l$ in
Eq. (\ref{sol2}) in general is a real number).  For the
correspondence between $\hat Y_{d-2}$ and $Y_{d-2}$ to be one-to-one,
we need to restrict $l$ in Eq.  (\ref{sol2}) to be also a nonnegative
integer.  If this is the case, both equations (\ref{sol2}) and
(\ref{sol2a}) can be solved in terms of ultraspherical (Gegenbauer)
polynomials.  Using the explicit form of $\hat Y_{d-2}$ and $Y_{d-2}$
one can then show that $Y_{d-2}$ is the analytical continuation
$\theta\to i \theta$ of $\hat Y_{d-2}$.

Let us now discuss the solutions of the radial equation (\ref{sol3}).
The solution of this differential equation can be expressed in terms
of hypergeometric functions
\beq \lb {hyp1} f(r)= r^lF\left( {1\over
2}(l+h_+),\,{1\over 2}(l+h_-),\,1+{n\over 2},\, 1+\laq r^2 \right),
\feq where $h_\pm$ are given by Eq. (\ref{e1}).

In the following sections we will need the asymptotical 
behavior $r\to\infty$ of the scalar field
$\phi$.
This can be easily read from Eq. (\ref{hyp1}), taking into account 
that  $h_+>0$ and $h_-<0$  we have 
\beq\lb{ab}
\phi\sim r^{-h_{-}}h(t)\hat 
Y_{d-1}(\Omega)=r^{-h_{-}}\phi_{-}(t,\Omega).
\feq
Notice that although the wave equation allows two possible behaviors 
at infinity, $\phi\sim r^{-h_{\pm}•}$, the falloff $\phi\sim r^{-h_{+}•}$
is always subleading with respect to  $\phi\sim r^{-h_{-}}$, as long as
we use the hyperbolic slicing (\ref{dsme}) where only the boundary
$\cal{I^{+}}$ is visible. The absence of the subleading behavior 
becomes a boundary condition, one can consistently impose on $\phi$.
This corresponds to appropriate boundary conditions for the Green functions 
(see Sect. 6). 
The previous discussion does not apply if we take dS$_{d}$ in the spherical 
slicing (\ref{gdsme}), in which both boundaries $\cal{I^{\pm}}$ are 
visible. For consistency we have now to keep both behaviors for $\phi$. 
The subleading behavior  in $\cal{I^{+}}$ becomes leading in
$\cal{I^{-}}$ and vice versa.  
The possibility of ruling out consistently, 
the falloff behavior $\phi\sim r^{-h_{+}}$ for the 
scalar field in  dS$_d$ 
is an important consistency check of our analytic continuation from 
dS$_d$ to AdS$_{d}$. For a scalar field in AdS$_{d}$  only one 
falloff for $\phi$ appears in the Green functions.

So far we  have considered a solution for $\phi$ in AdS$_{d}$ 
with global coordinates (\ref{adsme}) as analytic continuation of a solution 
in dS$_d$ in hyperbolic coordinated (\ref{dsme}).
We can also consider a solution for $\phi$  in AdS$_{d}$ in Poincar\'e 
coordinates (\ref{me}) as analytic continuation of a solution in  
dS$_d$ in the planar slicing (\ref{me1}).

The solution of the wave equation in the background metric (\ref{me1}) is
\beq\lb{sol4}
\phi_{dS}•=A e^{ik_{a}x_{a}} u^{(1-d)/2}Z_{\nu}(ku^{-1}),
\feq
where A is an integration constant, $k_{a}$ are real
(we consider for simplicity 
only oscillatory solutions on  $\cal{I^{+}}$),
$k^{2}=\sum_{a=1}^{d-1}k_{a}•^{2}$, $u=e^{\la \hat t}•$ and 
$Z_{\nu}$ are Bessel function 
with $\nu=(1/2)\sqrt{(d-1)^{2}+{4m^{2}/ \laq}}$.

Analytically continuing $x_{j}\to ix_{j},\, j=1\ldots(d-2)$, 
the solution (\ref{sol4}), we get  a solution of the wave equation
in the AdS background metric (\ref{me})
\beq\lb{sol4a}
\phi_{AdS}=A e^{-k_{j}x^{j} +ik_{d-1}\tau } u^{(1-d)/2}Z_{\nu}(ku^{-1}).
\feq

The asymptotic behavior $\hat t\to\infty$ ($u\to \infty$) of solutions 
(\ref{sol4}), (\ref{sol4a}) is easily found to be
\beq\lb{solas}
\phi\sim e^{-\la \hat t h_{-}}\phi_{-}(x) = u^{-h_{-}}\phi_{-}(x,\tau).
\feq
\section{Green functions}
Our next task is to consider Green functions.  Our goal is to obtain
Green functions for a scalar with square-mass $m^{2}$ propagating in
AdS$_{d}$ as analytical continuation of a tachyon with square-mass 
$-m^{2}$
propagating in dS$_d$.  There are two main problems that we have to
face in order to achieve this goal. (a) we can construct various
$SO(1,d)$ invariant vacua for a scalar field in dS$_d$ but the
$SO(2,d-1)$ invariant vacuum for a scalar in AdS$_{d}$ is unique. (b)
because we are considering propagation of a tachyon in dS$_d$,
infrared divergent terms are expected to appear in the correlation
functions.

 $SO(1,d)$-invariant Green functions on dS$_d$ have the general 
form \cite{Allen:ux}
\beq\lb{gf}
G(x,y)= F_{1}(P)\Theta(x,y)+ F_{2}(P)\Theta(y,x),
\feq
where $P$ depends on the geodetical distance $D(x,y)$ between the points 
$(x,y)$, $P=\cos{\la D(x,y)}$ and $\Theta(x,y)=(0,1/2,1)$ 
respectively for $x$ being in the (past of, spacelike separated from, 
future of) $y$.

In principle one could also consider more general $O(1,d)$-invariant 
Green functions. The non-connected group contains also the inversion, 
which in our pa\-ra\-me\-tri\-za\-tion (\ref{dsme}) acts as $r\to -r$. 
Because our coordinate system covers only the region $r> 0$ we have to 
exclude this case. 
It is also evident that for a field theory  defined on dS$_d$
parametrized as in Eq. (\ref{dsme}) $F_{2}=0$.
This follows from the discussion of the causal structure  of the
spacetime of Sect. 4. Our coordinate system covers only the future 
light-cone of a observer sitting at the equator of the $S^{d-1}$ sphere.
Owing to the 
spherical symmetry there is nothing particular about the equator, so 
the previous feature must hold for every point of the sphere.

A quantum field theory on dS$_{d}$  allows for a family of
$SO(1,d)$-invariant vacua \cite{Allen:ux,Bousso:2001mw}.  This
vacuum degeneracy is essentially due to a peculiarity of the causal
structure of  dS$_d$, which implies that we can have Green
functions with singularities not only when $x$ is in the light-cone of
$y$ ($P=1$) but also when $x$ is in the light cone of the point $\hat
y$ antipodal to $y$ ($P=-1$) \cite{Allen:ux}.  Of particular relevance
is the so-called Euclidean vacuum, which can be defined as the vacuum
in which the Green function is singular only if $x$ is in the light
cone of $y$, i.e for $P=1$.  These are features of a field theory
defined over the whole dS$_d$.  From the general theory of field
quantization on curved spaces, we know that a particular
coordinatization of the space, which may cover only a region of the
entire spacetime, singles out a particular vacuum for the field.  We can
now ask ourselves, which is the vacuum singled out by the
coordinatization (\ref{dsme}). It is not difficult to realize that
this is exactly the euclidean vacuum (or a thermalization of it).
This follows essentially from the fact that our coordinate system
covers only the future light-cone of $x$. We cannot see the
singularity for $P=-1$.

Let us now consider analytic continuation of the Green functions 
(\ref{gf}). Because the analytical continuation (\ref{ac}) changes the
signature of the embedding space from $(-1,1\ldots 1)$ to 
$(1,1,-1\ldots -1)$ (see equation (\ref{embe}), it  switches  
from $SO(1,d)$-invariant Green functions for a tachyon on dS$_d$
to  $SO(2,d-1)$-invariant Green functions for a  scalar 
particle with positive square mass  in AdS$_{d}$.
This is evident by considering, for instance, the Wightman function 
$G(x,y)=<0|\{\phi(x),\phi(y)\}|0>$
for a tachyon in  dS$_d$,
\beq\lb{wi}
\left(\nabla^{2}_{dS}+m^{2}\right)G(x,y)=0.
\feq
The analytical continuation (\ref{ac}) maps this equation into that
for a   scalar 
particle  in AdS$_{d}$, 
\beq\lb{wi1}
\left(\nabla^{2}_{AdS}-m^{2}\right)G(x,y)=0.
\feq

Summarizing, the Green functions for a scalar field in AdS$_{d}$ can be 
obtained as analytical continuation of the Green function for a 
tachyon in dS$_d$. The unique AdS-vacuum is obtained as analytical 
continuation of the tachyonic dS Euclidean vacuum, singled out by 
the parametrization of dS$_{d}$ (\ref{dsme}).

Let us now solve equation (\ref{wi}).
By setting $z=(1+P)/2$ equation (\ref{wi}) takes the form \cite{Candelas:du}
\beq\lb{hyper1}
z(1-z){d^{2}G\over dz^{2}}+\left({d\over 2}-dz\right){dG\over dz}
+ {m^{2}\over \laq}G=0.
\feq
The solution of this differential equation can be written in terms of 
the hypergeometric function $F(h_{+}, h_{-}, {d\over2},z)$, with 
$h_{\pm}$ given by Eq. (\ref{e1}). $F$ has a singularity for $z=1$, 
and, being $h_{-}<0$ also at $z=\infty$.
The singularity at $z=1$ is the usual short-distance singularity.
Conversely, the presence of the $z=\infty$ singularity is related to the 
tachyonic nature of the particle and implies the emergence of 
correlations that diverge at large distances. 
This behavior is rather unphysical, for instance implies violation 
of unitarity for the dual CFT living on the boundary of dS$_d$.
To solve the problem we will look for solutions of Eq. (\ref{hyper1}) 
that are regular for $z=\infty$.
For $h_{+}-h_{-}\neq\,$integer the general solution of the 
hypergeometric equation (\ref{hyper1}) can be written as
\eqn\lb{hypsol}
G(z)&=&C_{1}G_{1}+ C_{2}G_{2}
=Re\left\{ C_{1}(-z)^{-h_{+}}F
\left(h_{+}, h_{+}+1-{d\over 2}, 
h_{+}-h_{-}+1, {1\over z}\right)\right.\nonumber\\
&+& \left. C_{2} (-z)^{-h_{-}}
F\left(h_{-}, h_{-}+1-{d\over 2}, 
h_{-}-h_{+}+1, {1\over z}\right)\right\},
\feqn
where $C_{1,2}$ are integration constants. 
keeping in mind that $h_{+}$ is positive but $h_{-}$ is negative 
and that $F(\alpha,\beta,\gamma,0)=1$, regularity of the
solution at $z=\infty$ requires $C_{2}=0$.
The asymptotical $z\to\infty$  behavior of the solution is,
\beq\lb{hypsolas}
G(z)\sim C_{1} (-z)^{-h_{+}}.
\feq
Let us now consider  $h_{+}-h_{-}=\,$integer. This condition implies 
a discrete tachyon spectrum
\beq\lb{qt}
m^{2}= \laq \left[n^{2}+n(d-1)\right],
\feq
where $n$ is a positive integer. The conformal weights are also 
integer, from Eqs. (\ref{e1}),(\ref{qt}) it follows, $h_{+}=n+(d-1)$,\, 
$h_{-}=-n$. The solution of the differential Eq. (\ref{hyper1}) has again the 
form
$G(z)=C_{1}G_{1}+ C_{2}G_{2}$, with $G_{1}$ given as in Eq. 
(\ref{hypsol}). $G_{2}$ has a complicated expression  containing the 
hypergeometric function $F$, $\ln z$ and  power series, and  
diverges for $z\to \infty$. Again the regularity condition at 
$z=\infty$ requires $C_{2}=0$, so that the solution has the same form 
as in the previous case.

The integration constant $C_{1}$ can be fixed by requiring that the 
solution has the universal, $P=1$, ($D=0$), short-distance behavior
\cite{Spradlin:2001pw}
\beq\lb{sd}
G\sim{\Gamma({d\over 2})\over 2 (d-2)\pi^{d-2}} (D^{2})^{1-{d\over 2}}.
\feq

To complete our calculation of the Green functions on dS$_d$ and to 
perform the analytic continuation (\ref{ac}) to the Green function in 
AdS$_{d}$ we need to know explicitly the dependence of $z=(1+P)/2$ from the 
coordinates $(r,t,\theta, \theta_{k})$ of Eq. (\ref{dsme}).
Using the definition  of $P$ in terms of the embedding coordinates
$P=\laq X^{A}\eta_{AB}X'^{B}$ and Eq. (\ref{para}), one finds after 
some algebra
\eqn\lb{p}
P_{dS}&=&\laq rr' \left[ -\cosh\theta \cosh\theta' 
+\sinh\theta\sin\theta' \cos W_{d-3}(\Omega,\Omega')\right]\nonumber\\
&+&\sqrt{\left(\laq r^{2}+1\right)\left(\laq r'^{2} 
+1\right)}\cos \la (t-t'),
\feqn
where $W_{d-3}(\Omega,\Omega')$ is the geodesic distance between points of 
coordinates $\Omega=\,$   $(\theta_{1},\ldots\theta_{d-3})$, 
$\Omega=(\theta'_{1},\ldots\theta'_{d-3})$ on the $S^{d-3}$ sphere.
The asymptotic behavior of $P$ on $\cal{I^{+}}$ is
\beq\lb{Pinf}
\lim _{rr'\to \infty} P_{dS}= -\laq rr' \left[ \cosh\theta \cosh\theta' 
-\sinh\theta\sin\theta' \cos W_{d-3}(\Omega,\Omega')-\cos\la(t-t')\right].
\feq
We can now perform the analytic continuation $\theta\to i \theta$ in Eq. 
(\ref{p}) to get the corresponding expressions for (negative) AdS$_{d}$.
We have
\beq\lb{p1}
P_{AdS}=-\laq rr' \left[  \cos W_{d-2}•(\Omega,\Omega')
\right]
+\sqrt{\left(\laq r^{2}+1\right)\left(\laq r'^{2} 
+1\right)}\cos \la (t-t'),
\feq
where now $W_{d-2}(\Omega,\Omega')$ is the geodesic distance between points  
on the $S^{d-2}$ sphere.
Asymptotically we have,
\beq\lb{Pinf1}
\lim _{r,r'\to \infty} P_{AdS}= -\laq rr'\left[ \cos W_{d-2}
(\Omega,\Omega')-\cos \la (t-t')\right].
\feq
The minus sign in front of the expression  on the right hand of Eq. 
(\ref{p1}) and (\ref{Pinf1}) is due to the fact that the analytically 
continued spacetime is ``negative'' AdS$_{d}$. It is a direct consequence 
of the signature (\ref{embe}) of the embedding space.

Analogous calculations enable us to write $P_{ds}$ and $P_{AdS}$ in 
in the planar, respectively,  Poincar\'e coordinates of Eqs. (\ref{me1}),
(\ref{me}). We have
\eqn\lb{p2}
P_{dS}&=&\cosh\la (\hat t-\hat t')-{1\over 2}e^{\la (\hat t+\hat t')}
\laq |x-x'|^{2}, \nonumber\\
P_{AdS}&=&{1\over 2}\left[{u'\over u}+{u\over u'}+\laq uu'
\left(-(\tau-\tau')^{2}+  |x-x'|^{2}\right)\right],
\feqn
where $|x-x'|^{2}= \delta_{ab}(x^{a}-x'^{a})(x^{b}-x'^{b})$,  
$a,b=1\ldots(d-1)$ in the case of $P_{dS}$ and $a,b=1\ldots(d-2)$ in 
the case of $P_{AdS}$.
The corresponding asymptotical expressions are,

\eqn\lb{p3}
\lim _{\hat t,\hat t'\to \infty}P_{dS}&=&-{1\over 2}e^{\la (\hat 
t+\hat t')}\laq |x-x'|^{2},\\
\nonumber
\lim _{u,u\to \infty}P_{AdS}&=&{1\over 2}uu'\laq
\left(-(\tau-\tau')^{2}+ |x-x'|^{2}\right).
\feqn

\section{Analytical continuation from dS/CFT to AdS/CFT}
We will now discuss the correlation functions induced on the boundary 
of dS$_d$ by the propagation of the scalar field $\phi$ on the bulk
and their analytic continuation.
Following  Strominger \cite{Strominger:2001pn}, the two-point
correlator of an operator $\cal O_{\phi}$ on the boundary $\cal I^{+}$ of 
dS$_d$  is derived from the expression
\beq\lb{int}
{\cal J}=\lim_{r\to\infty}\int_{\cal I^{+}}dV_{d-1}dV'_{d-1}
\left\{(rr')^{d}\left[\phi(y)
\stackrel{\leftrightarrow}{\partial}_{r}G(r,y,r'y')
\stackrel{\leftrightarrow}{\partial}_{r'}\Phi(r',y')\right]\right\}_{r=r'}\,,
\feq
where $G$ is the de Sitter invariant Green function and $dV_{d-1}$ 
and $y=(\theta,t,\Omega)$ denote, respectively, the 
measure and the coordinates  of the boundary  $\cal{I^{+}}$.
For a tachyon in dS$_d$, we can find the boundary correlators 
using
Eqs.\ (\ref{ab}), (\ref{hypsolas}) and (\ref{Pinf}) in Eq. (\ref{int}). 
One easily 
finds the coefficient of the quadratic term in Eq.
(\ref{int}), which is  identified with  the two-point correlator of an 
operator ${\cal O}_{\phi}$ dual to $\phi_{-}$,
\beq\lb{corr}
\langle{\cal O}_{\phi}(y){\cal O}_{\phi}(y')\rangle_{dS}= {\ka_0 \over 
\left[\cosh\theta \cosh\theta'-\sinh\theta \sinh\theta' 
\cos W_{d-3}(\Omega, \Omega')- \cos{\la (t-t')}\right]^{h_{+}}}\,
\feq
where $\ka_0$ is a constant. 
We will show later in detail that the short-distance behavior of 
the correlation functions (\ref{corr}) is that pertaining to  a {\sl 
Euclidean} CFT 
operator with conformal weight $h_{+}$.

Performing the analytic continuation $\theta \to i \theta$ one finds 
the  correlation functions induced on the boundary $\cal{B}$ of AdS$_{d}$
by propagation of $\phi$ on the AdS bulk,
\beq\lb{corr1}
\langle{\cal O}_{\phi}(y){\cal O}_{\phi}(y')\rangle_{AdS}= {\ka_0 \over 
\left[ \cos W_{d-2}(\Omega, \Omega')-\cos{\la (t-t')}\right]^{h_{+}}}.
\feq
At short-distance, the  correlators describe  a 
$(d-1)$-dimensional  CFT in {\sl Minkowski} space.

The physical meaning of the correlation functions (\ref{corr}), 
(\ref{corr1}) and of the analytic continuation relating them, can be 
better understood considering the  simplest, $d=3$, case.
For $d=3$, Eq. (\ref{corr}) becomes
\beq\lb{corr2}
\langle{\cal O}_{\phi}(\theta,r){\cal O}_{\phi}(\theta', 
r')\rangle_{dS}= {\ka_0 \over 
\left[\cosh\Delta\theta-\cos{\Delta \la t}\right]^{h_{+}}}\,
\feq
where $\Delta\theta=\theta-\theta'$ and $\Delta t=t-t'$. 
Correlation functions of this kind have been already discussed in 
\cite{Klemm:2001ea}.
It  was argued that they are thermal correlation functions for a 
two-dimensional 
Euclidean CFT, 
with the compact dimension of the cylindrical geometry being  
spacelike rather then timelike (remember that in Eq. (\ref{corr2}) 
both $\theta$ and $t$ are spacelike coordinates).
Alternatively, one can interpret (\ref{corr2}) as a thermal correlator
at imaginary temperature $iT= i\la /2\pi $. In fact using complex 
coordinates, $(w,\bar w)$ the correlator (\ref{corr2}) can be written 
\cite{Klemm:2001ea}
\beq\lb{corr3}
\langle{\cal O}_{\phi}(w,\bar w){\cal O}_{\phi}(w',\bar 
w')\rangle_{dS}\sim \left[\sinh(i\pi T\Delta w) \sinh(i\pi T\Delta 
\bar w)\right]^{-h_{+}}.
\feq

For $d=3$ the analytically continued correlator (\ref{corr1}) is,
\beq\lb{corr4}
\langle{\cal O}_{\phi}(\theta,t){\cal O}_{\phi}(\theta',t')\rangle_{AdS}= {\ka_0
\over 
\left[\cos\Delta\theta-\cos\Delta \la t\right]^{h_{+}}}\,
\feq
where $\Delta\theta=\theta-\theta'$ and now $0\le \theta\le 2\pi$ is the angular 
coordinate of the $S^{1}$ sphere. 
Defining light-cone coordinates $\la x^{+}=\theta+\la t$, 
$\la x^{-}=\la t-\theta$
Eq. (\ref{corr4}) takes the form of  a thermal  correlator  at 
temperature $T= \la /2\pi $ in two-dimensional Minkowski space,
\beq\lb{corr5}
\langle{\cal O}_{\phi}(x^{+},x^{-}){\cal O}_{\phi}(x'^{+},x'^{-})
\rangle_{AdS}\sim \left[\sin(\pi T\Delta x^{+}•) \sin(\pi T\Delta 
x^{-}•)\right]^{-h_{+}}.
\feq
This is what one expects to happen because  $t$ is a timelike,
periodic coordinate of the AdS spacetime with metric (\ref{adsme}).

The analytic continuation (\ref{ac}) relating dS$_{3}$ with AdS$_{3}$
maps the thermal Euclidean 2D CFT at imaginary temperature
(\ref{corr3}) living on the boundary of $dS_{3}$ into the thermal
Euclidean 2D CFT with same, but real, temperature (\ref{corr5}) living
on the boundary of $AdS_{3}$.

Until now our discussion of the boundary correlation functions was 
confined to the case in which dS$_d$ and AdS$_{d}$ are described 
by the metrics (\ref{dsme}), respectively, (\ref{adsme}).
The emergence of thermal correlation functions is related with
the presence a compact  direction $t$ (spacelike for dS$_{d}$ and timelike 
for AdS$_{d}$). This is a peculiarity of the parametrizations (\ref{dsme}), 
(\ref{adsme}), which is in particular relevant when one wants to 
describe the large-distance behavior of the correlation function.
If one wants to describe the short-distance behavior of the 
correlation functions, the planar  (Poincar\'e) coordinates used in  
Eq. (\ref{me1}) ((\ref{me})) for dS$_d$ (AdS$_{d}$) are more appropriate.
Using the parametrization (\ref{me1}) for dS$_d$, the asymptotic form 
for the Green function (\ref{hypsolas}), for the scalar $\phi$ 
(\ref{solas}) and for $P$ (\ref{p3}) in the 
integral (\ref{int}), one finds for the  boundary correlators

\beq\lb{corr6}
\langle{\cal O}_{\phi}(x){\cal O}_{\phi}(x')\rangle_{dS}= {\ka_0' \over 
\left|x-x'\right|^{2h_{+}}}.
\feq
In planar coordinates, correlators induced by a tachyon on the 
boundary of dS$_d$ have the usual 
short-distance behavior of a $d-1$-dimensional Euclidean CFT.
The operators ${\cal O}_{\phi}$ have  conformal dimension $h_{+}$.
The analytic continuation (\ref{ac1}), which maps dS$_d$ in planar 
coordinates into   AdS$_{d}$ in Poincar\'e coordinates brings the 
correlators (\ref{corr6}) into the form
\beq\lb{corr7}
\langle{\cal O}_{\phi}(t,x){\cal O}_{\phi}(t',x')\rangle_{AdS}= {\ka_0' \over 
\left[(\tau-\tau')^{2}-|x-x'|^{2}\right]^{h_{+}}}\,
\feq
which are the correlation functions for a Lorentzian $(d-1)$-dimensional CFT
living on the boundary of AdS$_{d}$.

The tachyonic nature of the scalar field propagating in  de Sitter 
space is essential to assure that the analytic continuation maps the 
Euclidean CFT on $\cal{I^{+}}$ into the Minkowskian CFT on  $\cal{B}$.
The tachyonic character of $\phi$ implies that only the weight 
$h_{+}$ is positive, whereas $h_{-}$ is negative. We can therefore 
use the ``unitarity'' condition  on $G$ discussed in Sect. 6 to 
rule out operators with conformal dimension $h_{-}$ from the 
Euclidean CFT living on the boundary on dS$_d$. This is exactly 
what we need, since we know that only operators with conformal 
dimension $h_{+}$ appear in the Lorentzian CFT living on the boundary
of AdS$_{d}$.

\section{Conclusion}
In this paper we have investigated the analytical continuation from 
$d$-dimensional Lorentzian de Sitter  to $d$-dimensional Lorentzian 
anti-de Sitter spacetime. The obstructions 
to perform this continuation can be removed by choosing a particular 
coordinatization of dS$_{d}$ (physically, the coordinate system used by an 
observer at the equator of the $S^{d-1}$ sphere) and considering 
tachyons in dS$_{d}$ in correspondence with particles with positive 
square-mass in AdS$_{d}$.
An important result of our paper is that the AdS/CFT duality can be obtained 
as analytical continuation of the dS/CFT duality. 
The correlations induced by tachyons on the boundary of dS$_{d}$ are in 
correspondence with the correlations induced by particles with 
positive square-mass in the boundary of AdS$_{d}$. 

Because we know much more about  AdS/CFT  then about dS/CFT,
the results of this  paper could be very useful to improve our 
understanding of the dS/CFT duality.  In particular,  the correspondence 
we have found between dS/CFT and AdS/CFT could help us to clarify the 
status of the dS spacetime in the string theory context.
Moreover, a feature of our analytic continuation may hold the
key for understanding the problems that dS$_d$ inherits from its 
nature of time-dependent gravitational background.
Our analytical continuation exchanges, modulo the overall sign of the 
metric, the timelike direction of dS$_d$ with the radial spacelike 
direction of AdS$_{d}$. At the same time a spacelike direction of 
the dS$_{d}$ boundary becomes a timelike direction of the AdS$_{d}$ boundary.
This feature gives a simple explanation of how conserved quantities, 
associated with {\sl{spacelike}} Killing vectors in dS$_d$ are mapped into 
conserved quantities, associated with {\sl{timelike}} Killing vectors 
in AdS$_{d}$. The dS$_{d}\to$AdS$_{d}$ analytical continuation can be therefore  used to 
give  a natural definition of the mass and entropy of
the dS spacetime (see Ref. 
\cite{Balasubramanian:2001nb,Ghezelbash:2001vs} and Ref. \cite{Cadoni:2002kz}
for a discussion of the 2D case.
 
The exchange of the timelike/spacelike nature of directions,  in passing 
from dS$_{d}$ to AdS$_{d}$ seems to be at the core of the relationship between 
dS/CFT and AdS/CFT. For instance, we have seen in section 7 that 
boundary thermal correlators, with correlations along spacelike 
directions, in the case of dS$_d$ become thermal correlators, with 
correlations along timelike directions, for  AdS$_{d}$.
One could also  use this feature to try to circumvent the arguments of Ref.
\cite{Dyson:2002nt} against  the dS/CFT correspondence.

\end{document}